# The Wood-Saxon proton optical potential for p-nuclei

*Sukhendu* Saha[1,2,*], *Dipali* Basak[1,2] and *Chinmay* Basu[1]

[1]Saha Institute of Nuclear Physics, 1/AF Bidhannagar, Kolkata, 700064, India
[2]Homi Bhabha National Institute, Anushaktinagar, Mumbai, 400094, India

**Abstract.** A phenomenological mass-energy dependent proton optical model potential has been computed for p-nuclei. The parameters of the Wood-Saxon optical potential are found to be a good fit for proton elastic scattering data involving p-nuclei and elements with mass numbers near p-nuclei (within the range of 74 < A < 148) at energies around the Coulomb barrier of the system. The elastic scattering data were meticulously fitted using the SFRESCO code, allowing for the calculation of the real and imaginary parts of the Wood-Saxon optical potential. To validate the model, experimental proton capture cross-sections for $^{106}$Cd and $^{113}$In near the Coulomb barrier were compared with results obtained using the TALYS-1.96 code, showing better agreement than the available global proton optical model potential.

## 1 Introduction

P-nuclei, which encompass around 30-35 stable, neutron-deficient nuclei with mass numbers ranging from 74 to 196, are unique in that they are not primarily formed through the s- or r-processes [1]. Their isotopic abundances are notably lower compared to other stable isotopes of the same element. In the extreme conditions of stellar explosions, a series of (γ,n) reactions on s- and r-seed nuclei leads to the synthesis of proton-rich, stable isotopes. This process results in an increase in the neutron separation energy while simultaneously decreasing the proton and alpha separation energy. Consequently, (γ,p) and (γ,α) reactions become important contributors to p-nuclei production [2].

Performing direct γ-disintegration reactions in the laboratory is challenging, leading researchers to study these processes through inverse reactions based on the principle of detailed balance. To investigate experimental charge particle capture reactions in alignment with theoretical predictions, researchers often employ the Hauser-Feshbach (HF) statistical model. Notably, the selection of the proton optical model potential serves as a critical input parameter for HF calculations.

Deriving the optical potential formally can be a challenging and imprecise task, especially when it comes to non-locality and complexity in solving the Schrödinger equation for the system. In practice, phenomenological Optical Model Potentials (OMPs) are commonly employed and adjusted to compare the experimental data [3]. These potentials are typically treated as local in nature, simplifying the mathematical formulation. Microscopic potentials are also tried to describe the physical processes involved.

Empirical potentials typically rely on functional forms defined by a limited number of parameters that are fine-tuned to achieve the best possible fit with experimental data. In this study, the optical potential has been defined as follows:

$$U_{opt} = V_c(r) + V(r) + iW(r) + V_{SO}(r)$$

Where, the first term corresponds to the Coulomb term. The second and third terms pertain to the nuclear potential, where the second term is the real volume term, $-Vf_v(r)$ and the third term is the imaginary term consist of imaginary surface term, $-W_s g_w(r)$ and imaginary volume term, $-W_v f_w(r)$. The fourth term accounts for the spin-orbit interaction potential, which arises from the non-zero spins of the projectile and target nuclei [3,4]. The Wood-Saxon (WS) form factor for the real and imaginary volume terms are,

$$f_i(r) = \frac{1}{1+e^{\frac{r-R_i}{a_i}}} \quad i = V, W$$

The imaginary surface term,

$$g_w(r) = c\frac{df_w(r)}{dr} = \frac{\exp\left[\frac{r-R_w}{a_w}\right]}{\left(1+\exp\left[\frac{(r-R_w)}{a_w}\right]\right)^2}$$

The Coulomb term is typically computed by considering the interaction of a point charge particle with a charge sphere of radius $R_c$,

$$V_c = \begin{cases} \left(\frac{3}{2}-\frac{r^2}{2R_c^2}\right)\frac{Z_p Z_t e^2}{2} & r \leq R_c \\ \frac{Z_p Z_t e^2}{2}, & r > 0 \end{cases}$$

The spin-orbit term,

$$V_{SO}(r) = -V_{SO}\left(\frac{\hbar}{m_\pi c}\right)^2 \frac{1}{r}\frac{df_{so}(r)}{dr}\bar{l}.\bar{s}$$

The WS form factor for the spin-orbit term,

---

* Corresponding author: sukhendu.saha@saha.ac.in

$$f_{so}(r) = \frac{1}{1+e^{\frac{r-R_{so}}{a_{so}}}} \text{ and } \left(\frac{\hbar}{m_\pi c}\right)^2 \approx 2.00 \text{ fm}^2$$

Here, $m_\pi$ represents the pion mass, while $V$, $W_s$, $W_v$, and $V_{SO}$ denote the potential depths. $R_i$ and $a_i$ correspond to radii and diffusivity parameters, respectively [3].

In this work, The Wood-Saxon proton optical potential parameters were obtained by fitting them to the available proton elastic scattering data from the EXFOR database. This dataset encompassed 17 different nuclei, including p-nuclei and those within the mass range adjacent to p-nuclei (from 76 to 174). The proton energy considered was in the vicinity of the Coulomb barrier, ranging from 10 to 24 MeV. To verify these parameters, a comparison was made between the experimental proton capture cross-sections and theoretical predictions. The calculations involved the utilization of the Hauser-Feshbach statistical model to compute cross-sections using TALYS-1.96 [4].

## 2 Proton Optical Model Potential

Already available experimental proton elastic scattering data for 17 different elements/isotopes were collected (data retrieved from the EXFOR database on Jan 2023) and then fitted using the optical parameter search code SFRESCO [5]. Specifically, data from the lowest energy proton scattering within the range of 10 to 24.6 MeV were used for the calculations. The outcomes of the SFRESCO fitting process, including the chi-square ($\chi^2$) value, have been presented and are listed in **Table 1**.

Due to the low energy of the projectile particles, the imaginary volume potential term was omitted. This term accounts for the loss of projectile particles resulting from collisions with the nucleons of the target.

The $\chi^2$ value deteriorates notably for proton energies exceeding 20 MeV. The calculated potential was compared with both the Koning and Delaroche optical potential [6] and the Becchetti and Greenlees global optical potential parameters [7]. It was observed that within this specific energy and mass range, the calculated potential exhibited a more favourable fit to the elastic scattering data.

The experimental elastic scattering data reference has been tabulated in **Table 1**.

## 3 Results and discussion

The mass-energy dependent proton-optical potential, formulated in the Wood-Saxon form, is presented below,

$$V_v = 19.38 + 7.24 A^{\frac{1}{3}} - 0.43E + 40.1 \frac{N-Z}{A}$$
$$r_v = 1.01 \quad a_v = 0.67$$
$$W_s = 8.55 + 0.275 \frac{A}{N-Z}$$
$$r_i = 1.00 \quad a_i = 0.29 + 3.03 \frac{N-Z}{A}$$
$$V_{so} = 6.36 \quad r_i = 1.00 \quad a_{so} = 0.69$$

**Table 1.** The real and imaginary components of the proton optical potential, represented in the Wood-Saxon form, were determined for elements within the mass ranges of p-nuclei using the SFRESCO code. $E_{lab}$ denotes the proton energy in the laboratory frame. Parameters $V$, $W_s$, and $V_{so}$ represent the depth of the potential, while $r_v$, $r_i$, and $r_{so}$ denote the radii of the potential well. Parameters $a_v$, $a_i$, and $a_{so}$ correspond to the diffusivity parameters.

| Elements/ Isotopes | $E_{lab}$ (MeV) | Real volume terms | | | Imaginary surface terms | | | Real spin-orbit terms | | | $\chi^2/N$ | Ref |
|---|---|---|---|---|---|---|---|---|---|---|---|---|
| | | V | $r_v$ | $a_v$ | $W_s$ | $r_i$ | $a_i$ | $V_{so}$ | $r_{so}$ | $a_{so}$ | | |
| $^{76}$Se | 16 | 46.9 | 1.01 | 0.74 | 14.89 | 1 | 0.51 | 7.3 | 1 | 0.69 | 6.117 | [8] |
| | 22.3 | 42 | 1.05 | 0.63 | 15.4 | 1 | 0.5 | 6.59 | 1 | 0.35 | 67.8 | [9] |
| $^{86}$Sr | 24.6 | 45.31 | 1.04 | 0.78 | 12.61 | 1.02 | 0.5 | 5 | 1 | 0f.48 | 122.16 | [10] |
| $^{90}$Zr | 12 | 50.04 | 1.03 | 0.56 | 13.184 | 1 | 0.4 | 5.81 | 1.19 | 0.8 | 0.06 | [11] |
| | 16 | 48.96 | 1 | 0.63 | 7.69 | 1.05 | 0.66 | 7 | 1.03 | 0.67 | 13.5 | |
| $^{91}$Zr | 16 | 50.41 | 1 | 0.59 | 8.9 | 1.03 | 0.63 | 5.8 | 1.1 | 0.48 | 2.65 | [12] |
| $^{92}$Mo | 15 | 50.245 | 1 | 0.66 | 9.12 | 1 | 0.6 | 6.76 | 1 | 0.8 | 14.8 | [13] |
| $^{93}$Nb | 16 | 45.4 | 1.05 | 0.8 | 8.85 | 1.09 | 0.6 | 8 | 1 | 0.3 | 5.51 | [14] |
| $^{94}$Mo | 12.52 | 51.88 | 1 | 0.68 | 9.53 | 1 | 0.59 | 6.657 | 1.1 | 0.35 | 0.02 | [15] |
| $^{103}$Rh | 17 | 49.75 | 1 | 0.68 | 8.43 | 1.06 | 0.73 | 8 | 1 | 0.55 | 5.37 | [16] |
| $^{104}$Pd | 10.25 | 52.063 | 1 | 0.76 | 11.265 | 1.07 | 0.57 | 8.997 | 1 | 0.66 | 2.37 | [17] |
| | 12.1 | 52.154 | 1 | 0.698 | 10.211 | 1 | 0.66 | 5.76 | 1.28 | 0.79 | 2.59 | |
| | 15 | 51.354 | 1.01 | 0.658 | 8.13 | 1 | 0.78 | 7.2 | 1.01 | 0.8 | 2.73 | |
| $^{106}$Cd | 22.3 | 45.849 | 1.04 | 0.62 | 14.36 | 1 | 0.5 | 6.48 | 1.01 | 0.8 | 62.25 | [18] |
| $^{108}$Cd | 22.3 | 50.053 | 1 | 0.7 | 12.646 | 1 | 0.56 | 6.91 | 1 | 0.8 | 57.79 | |
| $^{112}$Sn | 20.51 | 43.388 | 1.09 | 0.616 | 14.352 | 1.02 | 0.5 | 6.78 | 1.01 | 0.49 | 2.82 | [19] |
| $^{116}$Sn | 16 | 50.362 | 1.01 | 0.69 | 9.26 | 1.04 | 0.67 | 6.72 | 1 | 0.35 | 19.07 | |
| $^{115}$In | 20.4 | 51.244 | 1 | 0.77 | 14.023 | 1.03 | 0.57 | 6.94 | 1 | 0.8 | 21.29 | [20] |
| $^{134}$Ba | 20.4 | 50.712 | 1.03 | 0.71 | 11.352 | 1.05 | 0.62 | 8 | 1.09 | 0.76 | 3.72 | [21] |
| $^{148}$Sm | 12 | 57.216 | 1 | 0.647 | 14.177 | 1 | 0.5 | 9 | 1.01 | 0.8 | 38.6 | [22] |
| | 16 | 49.8 | 1.04 | 0.77 | 7.94 | 1.15 | 0.76 | 8.24 | 1.01 | 0.35 | 5.58 | |
| $^{172}$Yb | 16 | 50.34 | 1.01 | 0.78 | 13.53 | 1 | 0.85 | 9 | 1.01 | 1 | 25.57 | [23] |

This potential has been derived from parameters obtained through the fitting of elastic scattering data.

The potential mentioned above involves only three dependent variables, making it relatively straightforward to fit with experimental data. In contrast, the Koning and Delaroche optical potential and the Becchetti and Greenlees global optical potential feature a larger number of dependent variables. Furthermore, the mentioned potential has demonstrated effective in achieving a satisfactory fit for the proton capture cross-sections of p-nuclei.

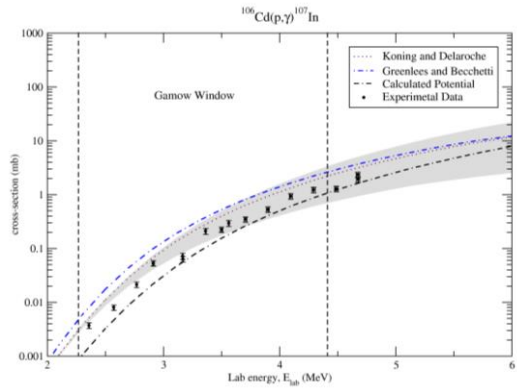

**Figure 1**

The proton capture cross-section data for $^{106}$Cd and $^{113}$In is depicted in Figure 1 and Figure 2. These experimental data points were fitted using TALYS-1.96 using different potential models. The shaded gray area in both figures represents the range of TALYS-1.96 predictions. The experimental data in **Figure 1** taken from [24] and expt data(1) and expt data(2) in **Figure 2** taken from [25].

Notably, the proton capture cross-sections for both elements were overestimated by the Greenlees and Becchetti and TALYS default potential models.

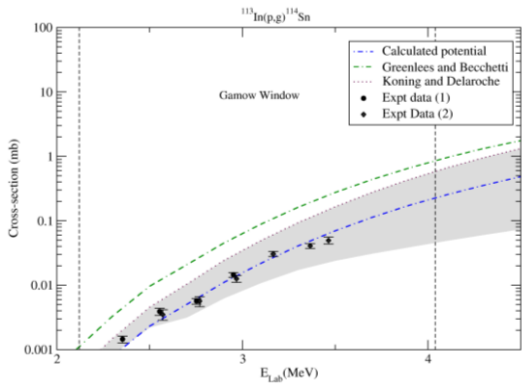

**Figure 2**

However, the potential calculated in this study provides reasonable estimates for the cross-section values.

## Acknowledgements

The authors express their gratitude to Dr. A.J. Koning for his valuable assistance during the theoretical calculations using the TALYS-1.96 code. Mr. Sukhendu Saha, affiliated with SINP, Kolkata, supported by the CSIR under file number 09/489(0119)/2019-EMR-I, acknowledges the funding provided by CSIR, India.